\documentstyle[twocolumn,aps,prd,epsf]{revtex}
\newcommand{\g}[1]{{\bf {#1}}}
\begin{document}
\draft
\twocolumn[\hsize\textwidth\columnwidth\hsize\csname @twocolumnfalse\endcsname

\title{Hydrodynamics of an ultra-relativistic fluid 
in the flat anisotropic cosmological model}
\author{V.P. Ruban
~ and ~D.I. Podolsky}
\address{L.D. Landau Institute for Theoretical Physics, 2 Kosygin St., 
117334, Moscow, Russia}
\date{February, 2001}
\maketitle

\begin{abstract}
Motion of an ultra-relativistic perfect fluid in space-time with 
the Kasner metrics is investigated by the Hamiltonian method. 
It is found that in the limit of small times a tendency takes 
place to formation of strong inhomogeneities in matter distribution. 
In the case of slow flows the effect of non-stationary anisotropy
on dynamics of sound waves and behaviour of frozen-in vortices is 
considered. It is shown that hydrodynamics of slow vortices on the 
static homogeneous background is equivalent to the usual Eulerian 
incompressible hydrodynamics, but in the presence of an
external non-stationary strain velocity field.
\end{abstract}

\vskip0.5pc]


Dynamics of a relativistic fluid in a gravitational field has a set
of distinctive properties in comparison with hydrodynamics in a flat
space-time \cite{LL6},\cite{LL2}. An explicit dependence of the metric 
tensor components $g_{ik}(t,\g{r})$ on time coordinate $t$ and spatial 
coordinates $\g{r}=(x,y,z)$, together with back influence of the 
matter on gravitational field, result in a wide variety in behaviour 
of solutions for equations of motion. Many phenomena in the 
general-relativistic hydrodynamics have not been studied yet. 
In order to get a better understanding of various processes taking 
place in the system, it is useful to analyse simple
limit cases, when only one of many general-relativistic effects is 
mostly important. If a dissipation may be neglected, then one can 
simplify significantly the analysis by using the Hamiltonian formalism,
application of which to hydrodynamics has given many interesting 
results (see the reviews \cite{ZK97}, \cite{M98} and references therein).
A brief discussion of adaptation of the Hamiltonian method for the
relativistic hydrodynamics can be found, for instance, in the papers
\cite{Brown} and \cite{Ruban2000PRD}. So, in the work 
\cite{Ruban2000PRD}, slow isentropic flows of an ideal relativistic
fluid have been considered and a partial study of influence of
a spatial inhomogeneity  on the motion of hydrodynamical
vortices has been performed. On the contrary, in present work we
suppose that the space is homogeneous and concentrate our attention on
the effects of its non-stationary anisotropy. It is known that there
exists the class of solutions of the Einstein equations for empty
space-time, which has been found by Kasner in 1922, with metrics
of the following form\cite{LL2}:
\begin{equation}\label{Kasner}
ds^2=dt^2-
t^{2/3}(t^{2\lambda_1}dx^2+t^{2\lambda_2}dy^2+t^{2\lambda_3}dz^2),
\end{equation}
where $\lambda_1, \lambda_2, \lambda_3$ are any three numbers 
satisfying two conditions
\begin{equation}\label{lambdas}
\lambda_1+\lambda_2+\lambda_3=0, \qquad 
\lambda_1^2+\lambda_2^2+\lambda_3^2={2}/{3}.
\end{equation}
At small times, this metrics describes approximately also the case of
Universe filled uniformly by a fluid, since in the limit
$t\to 0$ a back reaction of matter on the gravitational field
appears to be negligible \cite{LL2},\cite{BLKh1970}.
In present work we have used this result and considered the 
hydrodynamics of a perfect fluid having ultra-relativistic equation
of state, in the space-time with the fixed metrics (\ref{Kasner}).
We have established that at early times, flows with a finite mean 
momentum are in such a dynamic regime that a motion of each fluid 
element is nearly independent on the motion of other fluid elements. 
Due to this reason, there exists a tendency to formation of strong 
inhomogeneities in the spatial distribution of the matter
which is related to singularities of an approximate Lagrangian mapping.
We have found also that even in the case of slow flows with zero 
mean momentum the nonstationary anisotropy influences strongly both 
on the dynamics of sound oscillations and on the vortex motion,
making their properties be different in some aspects from properties 
of the analogous objects in the flat space-time.

The plan of further exposition is the following. At first, a necessary
explanation will be given concerning the employed theoretical method,
and the expression for the Lagrangian ${\cal L}$ of relativistic
ideal hydrodynamics in a given gravitational field will be written.
After a brief discussion of specific properties of corresponding
equations of motion, we shall pass to the Hamiltonian description
of the system dynamics and we will introduce the exact
expression for the Hamiltonian ${\cal H}$ corresponding to the 
ultra-relativistic equation of state of the matter. Some important 
properties of the dynamics in the limit $t\to 0$ will be discussed.
A quadratic (on small deviations) part ${\cal H}^{(2)}$ of the 
Hamiltonian will be used for the derivation of simplified
equations of motion for sound waves and vortices.


At first we introduce needed concepts. The energy density
$\varepsilon$, measured in the locally co-moving reference frame,
and the density $n$ of number of conserved particles are
connected by equation of state $\varepsilon=\varepsilon(n)$ \cite{LL6}.
The scalar $n$ is equal to the length of the 4-vector of current
$n^i=n(dx^i/ds)$ \cite{LL2}. Let us introduce the relative density
$\rho(t,\g{r})$ in such a manner that it obeys the standard
continuity equation
\begin{equation} \label{rho_t}
\rho_t+\nabla(\rho{\bf v})=0,
\end{equation}
where 3-velocity field $\g{v}(t,\g{r})=v^\alpha=(v^x,v^y,v^z)$ 
is defined as $dx^\alpha/dt$ on the world-line of the fluid point 
passing through $(t,\g{r})$. The equation (\ref{rho_t}) follows 
from the equation $n^i_{;i}=0$, expressing the conservation law 
for the fluid amount \cite{LL6}, if $\rho$ and $n$ are connected 
by the relation \cite{Ruban2000PRD}:
\begin{equation} \label{n_rho_g_v}
n=\frac{\rho}{\sqrt{|g|}}
\sqrt{g_{00}+2g_{0\alpha}v^\alpha+
g_{\alpha\beta}v^\alpha v^\beta},
\end{equation}
where $g=\mbox{det}\|g_{ik}\|$ is the determinant of the metric tensor.

The action functional $S=\int{\cal L}dt$ of relativistic
hydrodynamics is defined through the Lagrangian
${\cal L}={\cal L}\{\rho,\g{v}\}$ in the following way
\cite{Brown},\cite{Ruban2000PRD}:
\begin{equation} \label{Lagr_general}
{\cal L}\!=\!-\!\int\!\varepsilon\Big(\frac{\rho}{\sqrt{|g|}}
\sqrt{g_{00}+2g_{0\alpha}v^\alpha+
g_{\alpha\beta}v^\alpha v^\beta}\Big)\sqrt{|g|}d\g{r}.
\end{equation}

The equation of motion for the velocity field
$\g{v}(t,\g{r})$ has the structure
(generalized Euler's equation \cite{R99})
\begin{equation} \label{dynequation} 
(\partial_t+{\bf v\cdot\nabla})
\left(\frac{1}{\rho}\cdot\frac{\delta {\cal L}}{\delta {\bf v}}\right)=
\nabla\left(\frac{\delta {\cal L}}{\delta \rho}\right)-
\frac{1}{\rho}\left(\frac{\delta {\cal L}}{\delta v^\alpha}\right)
\nabla v^\alpha.
\end{equation}
This is merely the variational Euler-Lagrange equation expressing 
the least action $S$ principle for variations of world-lines of 
fluid particles. The 3-vector
$\g{p}(t,\g{r})=(1/{\rho})(\delta {\cal L}/{\delta {\bf v}})$,
\begin{equation} \label{p_def}
p_\alpha=\varepsilon'(n)\frac
{-(g_{0\alpha}+g_{\alpha\beta}v^\beta)}
{\sqrt{g_{00}+2g_{0\alpha}v^\alpha+g_{\alpha\beta}v^\alpha v^\beta}},
\end{equation}
is the canonical momentum of a liquid particle in the point
$(t,\g{r})$.
We see that the definition of the canonical momentum in relativistic
hydrodynamics depends in a complicated manner on the equation of state.
The relation
(\ref{p_def})  can be resolved respectively to the velocity,
which gives the inverse dependence: $\g{v}=\g{v}\{\rho,\g{p}\}$.
The remarkable equation for the vorticity field
$\g\Omega(\g{r},t)=\mbox{curl}\,\g{p}(t,\g{r})$
\begin{equation} \label{Omega_motion}
\g{\Omega}_t=\mbox{curl}[\g{v}\times\g{\Omega}],
\end{equation}
can be obtained from the equation (\ref{dynequation})
by applying the $\mbox{curl}$-operator. It implies that vorticity is
frozen in the fluid and the generalized Kelvin's theorem (conservation
of circulation of the field $\g{p}$ along arbitrary closed 
contour $\gamma(t)$ transported by flow) holds:
$$
\oint_{\gamma(t)}(\g{p}\cdot d\g{l})=\Gamma_\gamma=const.
$$

Passing to representation of flows in terms of the fields 
$\rho$ and $\g{p}$ and defining the Hamiltonian
 ${\cal H}\{\rho,\g{p}\}$ in such a manner:
\begin{equation}\label{Hdef}
{\cal H}\{\rho,\g{p}\}=
\Big(\int \Big(\frac{\delta{\cal L}}{\delta {\bf v}}
\cdot{\bf v}\Big)d{\bf r}-{\cal L}\Big)
\Big|_{\g{v}=\g{v}\{\rho,\g{p}\}}
\end{equation}
we have the equations of motion in the form
 \cite{R99}
\begin{equation}\label{drho/dt}
\rho_t+\nabla\left(\frac{\delta {\cal H}}{\delta {\bf p}}\right)=0,
\end{equation}
\begin{equation}\label{dp/dt}
{\bf p}_t=\left[\left(\frac{\delta {\cal H}}{\delta {\bf p}}\right)
\times\frac{\mbox{curl}\ {\bf p}}{\rho}\right]
-\nabla\left(\frac{\delta {\cal H}}{\delta \rho}\right).
\end{equation}

In particular, for the potential flows, when the vorticity is equal
to zero and
$\g{p}=\nabla\varphi$, the dynamical variables $\rho$ and $\varphi$
are canonically conjugated:
\begin{equation}\label{pair}
\rho_t=\frac{\delta{\cal H}\{\rho,\nabla\varphi\}}{\delta\varphi},\qquad
-\varphi_t=\frac{\delta{\cal H}\{\rho,\nabla\varphi\}}{\delta\rho}.
\end{equation}
Equations (\ref{pair}) will be used further (in linearized form)
for analysis of acoustic mode dynamics.

Other interesting dynamic regime can take place if there is an 
equilibrium solution $\rho=\rho_0(\g{r})$, $\g{p}=\g{p}_0=const$,
independent on time. If vortical component of flow is small and 
the sound waves are excited weakly, then perturbations of density
are negligibly small in comparison with its equilibrium value. 
In such circumstances a slow dynamics of the vorticity is described 
by the equation \cite{Ruban2000},\cite{Ruban2000PRD} 
\begin{equation}\label{Ham}
{\bf\Omega}_t=\mbox{curl}
\left[\mbox{curl}\left(\frac{\delta{\cal H}_*}{\delta{\bf\Omega}}\right)
\times\frac{{\bf\Omega}}{\rho_0(\g{r})}
\right],
\end{equation}
where the Hamiltonian of the vorticity
${\cal H}_*\{\g{\Omega}\}$ is obtained from the quadratic
(on $\delta\g{p}=\g{p}-\g{p}_0$ and
$\delta\rho=\rho-\rho_0$) part
${\cal H}^{(2)}\{\delta\rho,\delta\g{p}\}$ of the total Hamiltonian
after fixing the potential
$\varphi$ by condition of zero density variation 
$\delta{\cal H}^{(2)}/\delta\varphi=0$:
\begin{equation}\label{HamOmega}
{\cal H}_*\{\g{\Omega}\}=
{\cal H}^{(2)}\{0,\mbox{curl}^{-1}\g{\Omega}+\nabla\varphi\}
\Big|_{\delta{\cal H}^{(2)}/\delta\varphi=0}
\end{equation}
It should be noted that although the equilibrium solution
must be independent on time, the Hamiltonian itself
may contain an explicit dependence on $t$.
It is the case in the space-time with the Kasner metrics
(\ref{Kasner}).
In such a Universe the homogeneous equilibrium solutions
$\rho=\rho_0=const$, $\g{p}=\g{p}_0=const$ are possible\cite{LL2},
which fact can easily be seen from equations of motion
(\ref{rho_t}) and (\ref{dynequation}).

\begin{figure}
\epsfxsize=3.5in
\epsfysize=2.5in
\centerline{\epsffile{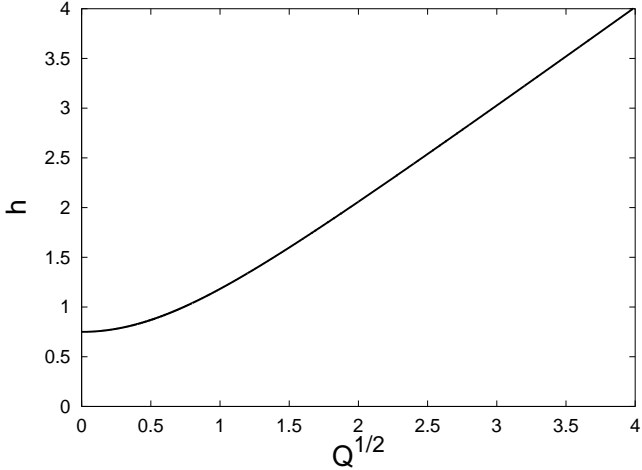}}
\caption{\small The dependence $h$ on $\sqrt{Q}$.}
\label{h_plot}
\end{figure}

Now we pass to concrete calculations. We will use the Kasner metrics
(\ref{Kasner}) and the ultra-relativistic equation of state
\begin{equation}\label{state_ultra}
\varepsilon(n)\sim n^{4/3},
\end{equation}
since $n\to\infty$ when $t\to 0$
\cite{LL2}.
For convenience, we change definition of the exponents:
$3\lambda_\alpha=2\mu_\alpha$ and perform the substitution
$t^{2/3}=\tau$ for the time coordinate in the action $S$,
and also choose in an appropriate manner scales for spatial
coordinates and for relative density
$\rho$,
in order to obtain
\begin{equation}\label{L_Kasner_ultra}
{\cal L}\{\rho,\g{v}\}=-\frac{3}{4}\int\rho^{4/3}
\Big(
1-\sum_\alpha\tau^{2\mu_\alpha}(v^\alpha)^2\Big)^{2/3}d\g{r}.
\end{equation}
The procedure of calculation of the Hamiltonian
${\cal H}\{\rho,\g{p}\}$, corresponding to this Lagrangian, includes
first of all the solution of the relation
$\g{p}=(1/{\rho})(\delta {\cal L}/{\delta {\bf v}})$, i.e.
\begin{equation}\label{P_ultra}
p_\alpha=\rho^{1/3}\Big(
1-\sum_\beta\tau^{2\mu_\beta}(v^{\beta})^2\Big)^{-\frac{1}{3}}
\tau^{2\mu_\alpha}v^{\alpha},
\end{equation}
respectively to $\g{v}$, and its substitution to (\ref{Hdef}).
Let us introduce the notations
\begin{equation}\label{QA_def}
A=\sum_\beta\tau^{2\mu_\beta}(v^{\beta})^2, \qquad
Q=\rho^{-2/3}\sum_\beta\tau^{-2\mu_\beta}p_{\beta}^2.
\end{equation}
We define  the auxiliary function $h(Q)$ by the conditions
\begin{equation}\label{h_def}
Q^3(1-A)^2=A^3, \qquad h=\sqrt{AQ}+\frac{3}{4}(1-A)^{2/3}.
\end{equation}
The explicit expression $h(Q)$ can be found with the help of
formula for roots of cubical polynome. Designating for brevity
\begin{equation}\label{RQ}
R(Q)=2^{-1/3}\left(27-2Q^3+3\sqrt{3}\sqrt{27-4Q^3}\right)^{1/3},
\end{equation}
we obtain that
$$
h(Q)=\frac{Q}{3}\left(-Q+\frac{Q^2}{R(Q)}+R(Q)\right)
$$
\begin{equation}\label{h_expression}
+\frac{3}{4}\Bigg(
1-\frac{Q}{9}
\left(-Q+\frac{Q^2}{R(Q)}+R(Q)\right)^2\Bigg)^{\frac{2}{3}}.
\end{equation}
We plot the dependence $h$ versus $\sqrt{Q}$ in the Fig. \ref{h_plot}.
At small $Q$, the approximate equality is valid
\begin{equation}\label{h_small_Q}
h(Q)\approx \frac{3}{4}+\frac{Q}{2}, \qquad Q\ll 1,
\end{equation}
and at $Q\gg 1$ the following behaviour takes place:
\begin{equation}\label{h_large_Q}
h(Q)\approx \sqrt{Q}+\frac{1}{4Q}, \qquad Q\gg 1.
\end{equation}

As the result, the Hamiltonian of the ultra-relativistic
fluid in the Kasner space-time has the form
\begin{equation}\label{H_Kasner_ultra}
{\cal H}\{\rho,\g{p}\}=\int\rho^{4/3}h\Big(\rho^{-2/3}
\sum_\alpha\tau^{-2\mu_\alpha}p_\alpha^2\Big)d\g{r}.
\end{equation}


For definiteness, we arrange the Kasner exponents in the order 
$\mu_1<\mu_2<\mu_3$. The constants $\mu_\alpha$ satisfy the inequalities
\begin{equation}
-2<2\mu_1<-1,\quad -1<2\mu_2<1,\quad 1<2\mu_3<2,
\end{equation}
following from the conditions (\ref{lambdas}).
Let us suppose $z$-component of the mean momentum be different from 
zero: $p_{z0}\not =0$. We consider here the case when deviations
of $\g{p}$ from the mean value are comparatively small.
Then at sufficiently small times the quantity
$Q$ reaches large values, since the order of magnitude of quantities 
$\rho$ and $\g{p}$ is conserved due to the conservation laws
for fluid amount and for mean momentum. Asymptotics
(\ref{h_large_Q}) gives the approximate Hamiltonian in this limit
\begin{equation} \label{H_appr}
\tilde{\cal H}
\approx\int\Big\{\rho\Big(
\sum_\alpha\tau^{-2\mu_\alpha}p_\alpha^2
\Big)^{\frac{1}{2}}
+\frac{\rho^2/4}{\Big(
\sum_\alpha\tau^{-2\mu_\alpha}p_\alpha^2
\Big)}\Big\}d\g{r}
\end{equation}
If we neglect here the quadratic on
$\rho$ term, where the coefficient in front of
$\rho^2$ is small in the limit of small times, then equation of
motion for the canonical momentum field does not depend on 
$\rho$ at all. Each point of fluid, labeled by a label
$\g{a}=a^\alpha$, moves almost independently on other points
in accordance with the approximate law
$$
x^\alpha(\tau,\g{a})\approx a^\alpha+\int_{0}^{\tau}
\frac{\tau^{-2\mu_\alpha}p_\alpha(\g{a})d\tau}
{\sqrt{\sum_\beta\tau^{-2\mu_\beta}p_\beta^2(\g{a})}}
$$
\begin{equation} \label{free_motion}
\approx a^\alpha+\frac{p_{\alpha}(\g{a})
\tau^{1+\mu_3-2\mu_\alpha}}
{|p_{z}(\g{a})|(1+\mu_3-2\mu_\alpha)}.
\end{equation}
Deviations of arbitrary constants of motion $\g{p}(\g{a})$ from 
the mean value make possible finite-time singularities of the mapping
$\g{r}(\tau,\g{a})$. Accordingly, quasi-two-dimensional domains of 
pancake-like shape may arise in the space where the density is much 
larger than the mean density: $\rho={\rho(\g{a})}
{\mbox{det}\|\partial\g{a}/\partial\g{r}\|}\gg\rho_0$.
There, the neglection of the quadratic on $\rho$
term is no more valid, and actually a redistribution
of momenta between fluid elements occurs.



To investigate slow flows on a resting background, it is
necessary to obtain the expression for
${\cal H}^{(2)}\{\delta\rho,\g{p}\}$. Let us expand
 ${\cal H}\{\rho,\g{p}\}$ to the second order terms on
$\delta\rho(\tau,\g{r})=\rho(\tau,\g{r})-\rho_0$ and $\g{p}(\tau,\g{r})$.
Then using of the expansion (\ref{h_small_Q}) gives
\begin{equation}\label{H2}
{\cal H}^{(2)}
=\frac{\rho_0^{2/3}}{2}
\int\sum_\alpha\tau^{-2\mu_\alpha}p_\alpha^2d\g{r}\,+\,
\frac{\rho_0^{-2/3}}{3}\int \frac{(\delta\rho)^2}{2} d\g{r}.
\end{equation}
Expression (\ref{H2}) and equations (\ref{pair}) allow us
immediately to write the linearized equations
of motion for the spatial Fourier-harmonics of sound waves:
\begin{equation}\label{sound}
-\frac{d\varphi_{\g{k}}}{d\tau}=
\frac{1}{3}\rho_0^{-\frac{2}{3}}\rho_{\g{k}},\quad
\frac{d\rho_{\g{k}}}{d\tau}=\rho_0^{\frac{2}{3}}\Big(
\sum_\alpha\tau^{-2\mu_\alpha}k_\alpha^2\Big)
\varphi_{\g{k}}.
\end{equation}
We can reduce this system to one ordinary linear differential equation
\begin{eqnarray}\label{sound2}
\frac{d^2\varphi_{\g{k}}(\tau)}{d\tau^2}+\frac{1}{3}\Big(
\sum_\alpha\tau^{-2\mu_\alpha}k_\alpha^2
\Big)\varphi_{\g{k}}(\tau)=0.
\end{eqnarray}
If $k_z^2\not =0$,
then behaviour of $\varphi_{\g{k}}(\tau)$ at small times
is determined by the exponent
$\mu_3$.
Obviously, all solutions of equation (\ref{sound2}) have in 
general an oscillating character at
$\tau>\sim\tau_*(\g{k})\sim|k_z|^{1/(\mu_3-1)}$.
There is no oscillations at $\tau\ll\tau_*(\g{k})$. 
Among linearly independent solutions there exists one, which tends 
to zero at $\tau\to 0$ as $\varphi^{(1)}_{\g{k}}\sim\tau$. 
Let us note that the corresponding $\rho^{(1)}_{\g{k}}$
tends to some finite value as $\tau\to 0$.
Another solution, linearly independent on first one, is finite
in the zero, but its derivative diverges at
$\tau\to 0$ as ${d\varphi^{(2)}_{\g{k}}}/{d\tau}\sim\tau^{1-2\mu_3}$,
and this fact means unbounded growth of $\rho^{(2)}_{\g{k}}$
and indicates transition of the system to the highly nonlinear 
dynamics regime described by the Hamiltonian (\ref{H_appr}).


For analysis of slow vortical flows, let us write explicitly the 
Hamiltonian ${\cal H}_*\{\g{\Omega}\}$. For simplicity we put 
$\rho_0=1$. In accordance with the definition
(\ref{HamOmega}) and the expression (\ref{H2}),
we obtain after simple calculations that
\begin{equation}\label{H_*}
{\cal H}_*\{\g{\Omega}\}
=\frac{1}{8\pi}\int\!\!\int
\frac{d\g{r}_1d\g{r}_2
\sum_\alpha \tau^{2\mu_\alpha}
\Omega^\alpha(\g{r}_1)\Omega^\alpha(\g{r}_2)}
{\Big(
\sum_\alpha\tau^{2\mu_\alpha}(x^\alpha_1-x^\alpha_2)^2\Big)^{1/2}}.
\end{equation}
Let us express the frozen-in field of the vorticity
through the shape
$\g{R}(\nu,\xi,\tau)$ of vortex lines,
(see \cite{R99}, \cite{Ruban2000}, \cite{KR2000PRE} for details):
\begin{equation}\label{lines}
{\bf \Omega }({\bf r},\tau)=\int_{\cal N}d^2\nu \oint 
\delta ({\bf r}-{\bf R}(\nu,\xi,\tau))
\frac{\partial{\bf R}}{\partial\xi}d\xi,  
\end{equation}
where $\nu$ is a vortex line label, which belongs to the 2D
manifold ${\cal N}$, $\xi$ is a longitudinal parameter along the line.
The dynamics of vortex lines is determined by the variational
principle
$\delta S_{\g{R}}=\delta\int{\cal L}_{\g{R}}d\tau=0$, with
the Lagrangian of vortex lines having the following form
\cite{Ruban2000PRD},
\cite{R99}-\cite{KR2000PRE}:
\begin{equation}
{\cal L}_{\g{R}}=\frac{1}{3}
\int_{\cal N}d^2\nu\!\oint\! \Big(\left[ {\bf R}_{\tau}\times
{\bf R}\right]
\!\cdot\!
{\bf R}_\xi\Big)d\xi 
 - {\cal H}_*\{{\bf \Omega }\{{\bf R}\}\!\},  \label{LAGR_lines}
\end{equation}
where the subscripts $\tau$ and $\xi$ mean the corresponding partial
derivatives.
Performing the deformating substitution
\begin{equation}\label{deformation}
\g{R}=(X,Y,Z)=(\tau^{-\mu_1}\tilde X,\,\,
\tau^{-\mu_2}\tilde Y,\,\,\tau^{-\mu_3}\tilde Z),
\end{equation}
we obtain the Lagrangian for $\tilde\g{R}(\nu,\xi,\tau)$:
$$
{\cal L}_{\tilde\g{R}}=\frac{1}{3}
\int \Big(\left[\tilde {\bf R}_{\tau}\times
\tilde{\bf R}\right]
\cdot
\tilde{\bf R}_\xi\Big)d\xi d^2\nu 
$$
$$
-\frac{1}{8\pi}\int\!\!\int\frac
{(\tilde\g{R}_{\xi_1}(\nu_1,\xi_1)\cdot\tilde\g{R}_{\xi_2}(\nu_2,\xi_2))}
{|\tilde\g{R}(\nu_1,\xi_1)-\tilde\g{R}(\nu_2,\xi_2)|} 
d\xi_1 d^2\nu_1 d\xi_2 d^2\nu_2
$$
$$
-\frac{1}{3\tau}\int 
\Big(\tilde X_{\xi}\tilde Y \tilde Z(\mu_2-\mu_3)
+\tilde Y_{\xi}\tilde Z \tilde X(\mu_3-\mu_1)
$$
\begin{equation} \label{L_deform}
+\tilde Z_{\xi}\tilde X \tilde Y(\mu_1-\mu_2)
\Big)d\xi d^2\nu
\end{equation}
It is interesting to note that the same Lagrangian for vortex lines
corresponds to the usual Eulerian incompressible hydrodynamics in
isotropic space with coordinates 
$\tilde\g{r}=(\tilde x,\tilde y,\tilde z)=
(\tau^{\mu_1}x,\,\tau^{\mu_2}y,\,\tau^{\mu_3}z)$, but
in the presence of external straining potential velocity field
\begin{equation} \label{v_deform}
\tilde\g{v}_{ext}(\tau,\tilde\g{r})=
\tau^{-1}(\mu_1\tilde x,\,\,\mu_2\tilde y,\,\,\mu_3\tilde z).
\end{equation}

In conclusion, let us say a few words about hydrodynamics in 
the homogeneous cosmological models of the other types \cite{LL2}, 
where the space is not flat. In this case, a quantitative analysis 
is difficult due to curvature of space and a complicated dependence of
metrics on time, but the qualitative 
result about initial free motion regime for fluid elements remains 
valid.  The reason is that an appropriately defined relative
density $\rho$ is conserved in order of magnitude, as well as 
covariant components $p_\alpha$ of the momentum field, while 
contravariant components $p^\alpha$ tend to infinity at $t\to 0$, 
like in the flat model. This fact means an infinite growth of
the quantity $p_\alpha p^\alpha$ and approaching of the velocity 
magnitude $\sqrt{v_\alpha v^\alpha}$ to the speed of light, regardless 
field $\rho$. Therefore, a weak initial spatial inhomogeneity in 
distribution of $p_\alpha(\g{a})$ may result, after some time,
in a strong spatial inhomogeneity of $\rho$.

\medskip

These investigations were supported by the RFBR (grants 00-01-00929
and 99-02-16224),
by the Russian State Program of Support of the Leading Scientific 
Schools (grants 00-15-96007 and 00-15-96699),
and by the INTAS.

\end{document}